\documentclass[draft]{agujournal}
\usepackage[T1]{fontenc}
\usepackage[latin9]{inputenc}
\usepackage{xcolor}
\usepackage{pdfcolmk}
\usepackage{url}
\usepackage{amstext}
\usepackage{graphicx}
\PassOptionsToPackage{normalem}{ulem}
\usepackage{ulem}

\makeatletter

\providecolor{lyxadded}{rgb}{0,0,1}
\providecolor{lyxdeleted}{rgb}{1,0,0}

\DeclareRobustCommand{\lyxsout}[1]{\ifx\\#1\else\sout{#1}\fi}

\usepackage{lmodern}
\usepackage[T1]{fontenc}
\usepackage{bm}
\usepackage{amsfonts}

\makeatother

\begin{document}

\title{Discovering state-parameter mappings in subsurface models using generative
adversarial networks}

\authors{\authors{Alexander Y. Sun \affil{1}}}

\affiliation{1}{Bureau of Economic Geology, Jackson School of Geosciences, The University of Texas at Austin, Austin, TX}

\correspondingauthor{A. Y. Sun}{alex.sun@beg.utexas.edu}
\begin{abstract}
A fundamental problem in geophysical modeling is related to the identification
and approximation of causal structures among physical processes. However,
resolving the bidirectional mappings between physical parameters and
model state variables (i.e., solving the forward and inverse problems)
is challenging, especially when parameter dimensionality is high.
Deep learning has opened a new door toward knowledge representation
and complex pattern identification. In particular, the recently introduced
generative adversarial networks (GANs) hold strong promises in learning
cross-domain mappings for image translation. This study presents a
state-parameter identification GAN (SPID-GAN) for simultaneously learning
bidirectional mappings between a high-dimensional parameter space
and the corresponding model state space. SPID-GAN is demonstrated
using a series of representative problems from subsurface flow modeling.
Results show that SPID-GAN achieves satisfactory performance in identifying
the bidirectional state-parameter mappings, providing a new deep-learning-based,
knowledge representation paradigm for a wide array of complex geophysical
problems. 
\end{abstract}

\section*{Plain Language Summary}

Development of physically-based models requires two steps, mathematical
abstraction (forward modeling) and parameter estimation (inverse modeling).
A high-fidelity model requires high-quality parameter support. The
need for identifying forward and reverse mappings (i.e., a function
that associates element of one set to another) is thus ubiquitous
in geophysical research. A significant challenge in geosciences is
that geoparameters are spatially heterogeneous and high dimensional,
and yet can only be observed at limited locations. The conventional
workflow, built on minimizing the model-observation mismatch at measurement
locations, does not offer an efficient way for estimating the spatial
structure of high-dimensional parameter fields. This work presents
a deep-learning-based framework for identifying the state-parameter
bidirectional mappings using the recently introduced generative adversarial
networks (GANs). GANs have been shown to be adept at associating images
from one domain to another. Its potential for discovering mappings
in physically based models has not been demonstrated so far. This
work shows that GAN can achieve high performance in learning bidirectional
parameter-to-state mappings in physically based models, thus providing
a new way of thinking and doing things in geosciences. The implication
for additional applications in subsurface modeling is significant.

\section{Introduction}

Deep learning (DL) has achieved great success in image recognition
and business intelligence over the past decade, continuously narrowing
the gap between artificial intelligence and human intelligence. Tremendous
interests exist in the geophysical research community to leverage
the strength of DL for solving similar image recognition and prediction
problems, such as land cover and land use classification \citep{castelluccio2015land},
extreme weather event forecasting \citep{xingjian2015convolutional,liu2016application},
estimation of particulate matter levels \citep{li2017estimating},
and data imputation \citep{fang2017prolongation}. To achieve high
accuracy, many DL algorithms require a large amount of labeled training
data (i.e., co-observed predictors and predictands), which is generally
hard to acquire in geoscience domains due to the invisibility of subsurface
processes and sparsity of in situ monitoring networks. 

In machine learning, semi-supervised learning has been used to tackle
the issue of limited labeled data \citep{chapellesemisupervised}.
As its name suggests, semi-supervised learning sits in between the
traditional unsupervised learning (all training data are unlabeled)
and supervised learning (all training data are labeled). Semi-supervised
learning methods use unlabeled data, together with limited labeled
data, to build better machine learning models, under the assumption
that unlabeled data are more abundant and carry information that is
useful for the inference of target variables \citep{chapellesemisupervised}. 

Semi-supervised learning may help solve a fundamental problem in geosciences,
namely, estimating the underlying generative model of sampled data
so that new samples can be synthesized from the learned model. In
fact, generative process modeling is at the core of all physical sciences,
where mechanistic models have long been applied to extract, abstract,
and approximate the observed causal structures in order to simulate
samples of the underlying physical processes. In geosciences in particular,
parametric forward modelings are carried out by solving partial differential
equations (PDEs) governing the spatially and/or temporally varying
subsurface physical processes, whereas inverse problems are formulated
to identify the model parameters by using observations of state variables
\citep{sun2015model}. Ideally, both the forward and inverse modeling
should be done in a closed loop manner such that new information can
be continuously assimilated to reduce uncertainty. Thus, the need
for resolving the bidirectional mappings between state and parameter
spaces always exists. An outstanding challenge is that many subsurface
processes are nonlinear, multiscale, and high-dimensional, making
it nontrivial to establish such mappings in practice.

The recently introduced generative adversarial networks (GANs), which
may be considered a subclass of DL for semi-supervised learning, hold
strong promises not only for learning the generative processes of
high-dimensional images with limited labeled data, but also for translating
seemingly unrelated images across different domains \citep{goodfellow2014generative}.
An open question is whether these interesting features of GANs can
benefit the geophysical modeling community. Here I explore the bidirectional
mapping capability of GANs and hypothesize that GANs may provide a
new workflow for inferring parameter-state mappings. Many of the conventional
parameter estimation methods are built on the minimization of certain
distance measures between observed and simulated values by running
a forward model iteratively, while uncertainty quantification is usually
done a posteriori. Under the GAN framework, the model parameter space
and state space are regarded as two inherently related image domains,
and DL-based functional relationships are obtained to facilitate the
cross-domain learning, namely, estimating parameters for given model
states, and vice versa. 

The main purpose of this study is to formulate a state-parameter identification
GAN (SPID-GAN) for obtaining deep bidirectional representations of
geophysical models. In the following, I first introduce the proposed
SPID-GAN framework, which combines the traditional geostatistical
simulation, physically-based forward modeling, and the point-based
parameter estimation workflow with cross-domain deep learning. The
framework is demonstrated using two different examples from subsurface
flow modeling, in which the model parameters are spatially heterogeneous,
representing high-dimensional samples obtained from single- and bimodal
distributions. I show that the DL-based SPID-GAN is well adept at
learning the subtle spatial patterns in model states and parameter
fields, thus representing a powerful tool for approximating the causal
structures of physical models. 

\section{Methodology}

\subsection{SPID-GAN framework}

The original GAN introduced by \citet{goodfellow2014generative} consists
of a pair of discriminator and generator, designed to compete with
each other as in a two-player game, thus the word adversarial. The
role of the generator is to create ``fake'' samples that are indistinguishable
from the training data, while the role of the discriminator is to
classify the generated samples to determine whether they are real
or fake. Let $\mathcal{G}$ denote a generator model that defines
a mapping $\mathcal{G}:X\rightarrow Y$, namely, it takes ${\bf x}\in X$
as input and generates a fake sample $\mathcal{G}({\bf x})$ that
has the same support as the training data ${\bf y}\in Y$. In addition,
let $\mathcal{D}$ denote a discriminator that determines whether
a sample is drawn from the empirical distribution of training data
$p_{data}({\bf y})$, or from the generator distribution $p_{model}(\mathcal{G}({\bf x}))$.
The goal of the generator is thus to push its sample distribution
$p_{model}(\mathcal{G}({\bf x}))$ toward the data distribution $p_{data}({\bf y})$.
At optimality the discriminator is maximally confused and cannot distinguish
real samples from ones that are fake (i.e., predicting with a probability
of 0.5 for all inputs) \citep{goodfellow2016nips}. Note that similar
principles are behind many Bayesian statistical inversion and ensemble-based
data assimilation algorithms, in which the common goal is updating
a prior distribution to a posterior distribution that reflects the
newer information. In practice, many of the conventional methods are
either limited to low-dimensional problems (e.g., particle filter
and Markov chain Monte Carlo) or to multivariate Gaussian distributions
(e.g., the ensemble Kalman filter). In comparison, GANs make no such
assumptions on the distributions of domain data. 

Existing GANs differ by how domains are defined and what cross-domain
mappings are to be learned. The generator in the original GAN takes
a sample from a low-dimensional latent space (i.e., a random noise
vector) and turns it into a real image (e.g., a car). Such a generative
process resembles the PCA-based random field simulation algorithm
commonly used in geostatistics \citep{satija2015direct}. The main
difference is that GANs train DL models to learn complex structural
patterns embedded in the training data, while the eigenvector-based
representation in PCA is linear and restricted to the 2nd-order statistics.
Since the work of \citet{goodfellow2014generative}, a large number
of GAN models have been introduced for cross-domain learning. So far,
GANs have been demonstrated in a number of inspiring applications,
such as (a) image superresolution, where low-resolution images are
used to generate their high-resolution counterparts; examples include
deep convolution GAN (dcGAN) \citep{radford2015unsupervised} and
superresolution GAN (SRGAN) \citep{ledig2017photo}; (b) cross-domain
image-to-image translation, where labeled information in the form
of either text descriptions or images is used to generate images in
another domain; examples include the conditional GAN \citep{mirza2014conditional},
coupled GAN \citep{liu2016coupled}, DiscoGAN \citep{kim2017learning},
and DualGAN \citep{yi2017dualgan}. Newer GANs can perform direct
cross-domain learning without using low-dimensional latent space vectors
as done in the original GAN.

Building on the existing cross-domain learning GANs, SPID-GAN approaches
the state-parameter bidirectional mapping problem by using two pairs
of generators and discriminators 
\begin{equation}
\begin{array}{c}
\mathrm{\textrm{Forward mapping }}\mathcal{G}_{PS}:\mathbb{R}_{P}\rightarrow\mathbb{R}_{S},\;\mathcal{D}_{S}:\mathbb{R}_{S}\rightarrow[0,1],\\
\textrm{Reverse mapping }\mathcal{G}_{SP}:\mathbb{R}_{S}\rightarrow\mathbb{R}_{P},\;\mathcal{D}_{P}:\mathbb{R}_{P}\rightarrow[0,1].
\end{array}
\end{equation}

\noindent where $\mathcal{G}_{PS}$ defines a forward mapping from
the parameter space $P$ to the model state space $S$, while $\mathcal{G}_{SP}$
provides a reverse mapping from $S$ to $P$. The two discriminators
$\mathcal{D}_{S}$ and $\mathcal{D}_{P}$ are used to determine the
authenticity of samples generated for the respective domains in terms
of probability. A practical working assumption on $\mathcal{G}_{PS}$
and $\mathcal{G}_{SP}$ is that they are bijective, meaning each element
of the domain $P$ is mapped by exactly one element of the domain
$S$. This helps prevent the many-to-one mappings during training,
which is also known as the mode collapse problem \citep{zhu2017unpaired}.
Another assumption is the continuity of mappings, namely, if two elements
in domain $P$ are close, then also should be the corresponding elements
in domain $S$. The same assumption is also implied by the stability
requirement of inverse solutions \citep{sun2015model}. Thus, to arrive
at meaningful solutions, one needs not only a proper algorithm design
(detailed below), but also an appropriate experimental design (next
section). 

The loss function used for training $\mathcal{G}_{PS}$ consists of
three terms \citep{kim2017learning,zhu2017unpaired}
\begin{eqnarray}
J_{PS}^{(\mathcal{G})}\left(P,S,\theta^{(\mathcal{D}_{S})},\theta^{(\mathcal{G}_{PS})},\theta^{(\mathcal{G}_{SP})}\right) & = & J_{PS}^{(\mathcal{D}_{S})}+d_{tran}\left(\mathcal{G}_{PS}({\bf x},\theta^{(\mathcal{G}_{PS})}),{\bf y}\right)+\nonumber \\
 &  & d_{cyc}\left(\mathcal{G}_{SP}\left(\mathcal{G}_{PS}({\bf x},\theta^{(\mathcal{G}_{PS})}),\theta^{(\mathcal{G}_{SP})}\right),{\bf x}\right),\label{eq:2}
\end{eqnarray}

\noindent where ${\bf x}\in P$, ${\bf y}\in S$, and $\theta^{(\cdot)}$
denote the unknown parameters of the respective generators/discriminators.
Before training, a standard practice in DL is to scale all training
data to the same range (e.g., $[0,1]$) so the objective functions
are addable. The first term on the right-hand-side of Eq. (\ref{eq:2})
defines the discriminator loss in the sense of the original GAN, 
\begin{eqnarray}
J_{PS}^{(\mathcal{D}_{S})}\left(\theta^{(\mathcal{D}_{s})},\theta^{(\mathcal{G}_{PS})}\right) & = & \frac{1}{2}\mathbb{E}_{{\bf x}\sim p_{data}({\bf x})}\left\Vert \mathcal{D}_{S}\left(\mathcal{G}_{PS}\left(\mathbf{x},\theta^{(\mathcal{G}_{PS})}\right),\theta^{(\mathcal{D}_{s})}\right)\right\Vert _{2}^{2}+\nonumber \\
 &  & \frac{1}{2}\mathbb{E}_{{\bf y}\sim p_{data}({\bf y})}\left\Vert \mathcal{D}_{S}\left({\bf y},\theta^{(\mathcal{D}_{S})}\right)-1\right\Vert _{2}^{2},\label{eq:3}
\end{eqnarray}

\noindent for which the goal is to minimize the error rate of the
discriminator on the fake sample (toward 0) and on the real sample
(toward 1). In Eq. (\ref{eq:3}), the expectation ($\mathbb{E}$)
is taken over all training samples. Note that the mean square error
(MSE) is used here, instead of the binary entropy loss used by \citet{goodfellow2014generative}. 

The second term on the right-hand-side of Eq. (\ref{eq:2}) measures
the translation loss in terms of mean absolute error (MAE), 
\begin{equation}
d_{tran}\left(\mathcal{G}_{PS}({\bf x},\theta^{(\mathcal{G}_{PS})}),{\bf y}\right)=\mathbb{E}_{{\bf y}\sim p_{data}({\bf y})}\left|\mathcal{G}_{PS}\left(\mathbf{x},\theta^{(\mathcal{G}_{PS})}\right)-{\bf y}\right|,
\end{equation}

\noindent where expectation is calculated based on all pairs of generated
and training images. The last term on the right-hand-side of Eq. (\ref{eq:2})
quantifies the cycle consistency (or reconstruction loss) between
the two generators using MAE 
\begin{equation}
d_{cyc}\left(\mathcal{G}_{SP}\left(\mathcal{G}_{PS}({\bf x},\theta^{(\mathcal{G}_{PS})}),\theta^{(\mathcal{G}_{SP})}\right),{\bf x}\right)=\mathbb{E}_{{\bf x}\sim p_{data}({\bf x})}\left|\mathcal{G}_{SP}\left(\mathcal{G}_{PS}({\bf x},\theta^{(\mathcal{G}_{PS})}),\theta^{(\mathcal{G}_{SP})}\right)-{\bf x}\right|.
\end{equation}

\noindent By minimizing the reconstruction loss, the cycle consistency
term helps to mitigate the mode collapse problem \citep{kim2017learning,zhu2017unpaired}.
The loss function of the generator $\mathcal{G}_{SP}$ can be defined
similarly to Eq. (\ref{eq:2}) by switching $P$ and $S$. The total
generator loss function $J^{(\mathcal{G})}$ for SPID-GAN is the average
of the two generator losses. Each discriminator uses a loss function
in the same form as Eq. (\ref{eq:3}) but with the opposite sign.
The generator and discriminator loss functions are highly coupled
and need to be solved from the following minimax optimization problem
\begin{equation}
\hat{\theta}^{(\mathcal{G}_{PS})},\hat{\theta}^{(\mathcal{D}_{S})},\hat{\theta}^{(\mathcal{G}_{SP})},\hat{\theta}^{(\mathcal{D}_{P})}=\arg\min_{\mathcal{\theta}^{(\mathcal{G}_{PS})},\mathcal{\theta}^{(\mathcal{G}_{SP})}}\max_{\theta^{(\mathcal{D}_{S})},\theta^{(\mathcal{D}_{P})}}J^{(\mathcal{G})}\left(\theta^{(\mathcal{G}_{PS})},\theta^{(\mathcal{D}_{S})},\theta^{(\mathcal{G}_{SP})},\theta^{(\mathcal{D}_{P})}\right).\label{eq:6}
\end{equation}

\noindent In practice, the optimization problem in Eq. (\ref{eq:6})
is solved by using alternating gradient updating steps for generators
and discriminators, with parameters of one group fixed when parameters
of the other are being updated in each iteration.

\begin{figure}
\noindent \centering{}\includegraphics[scale=0.8]{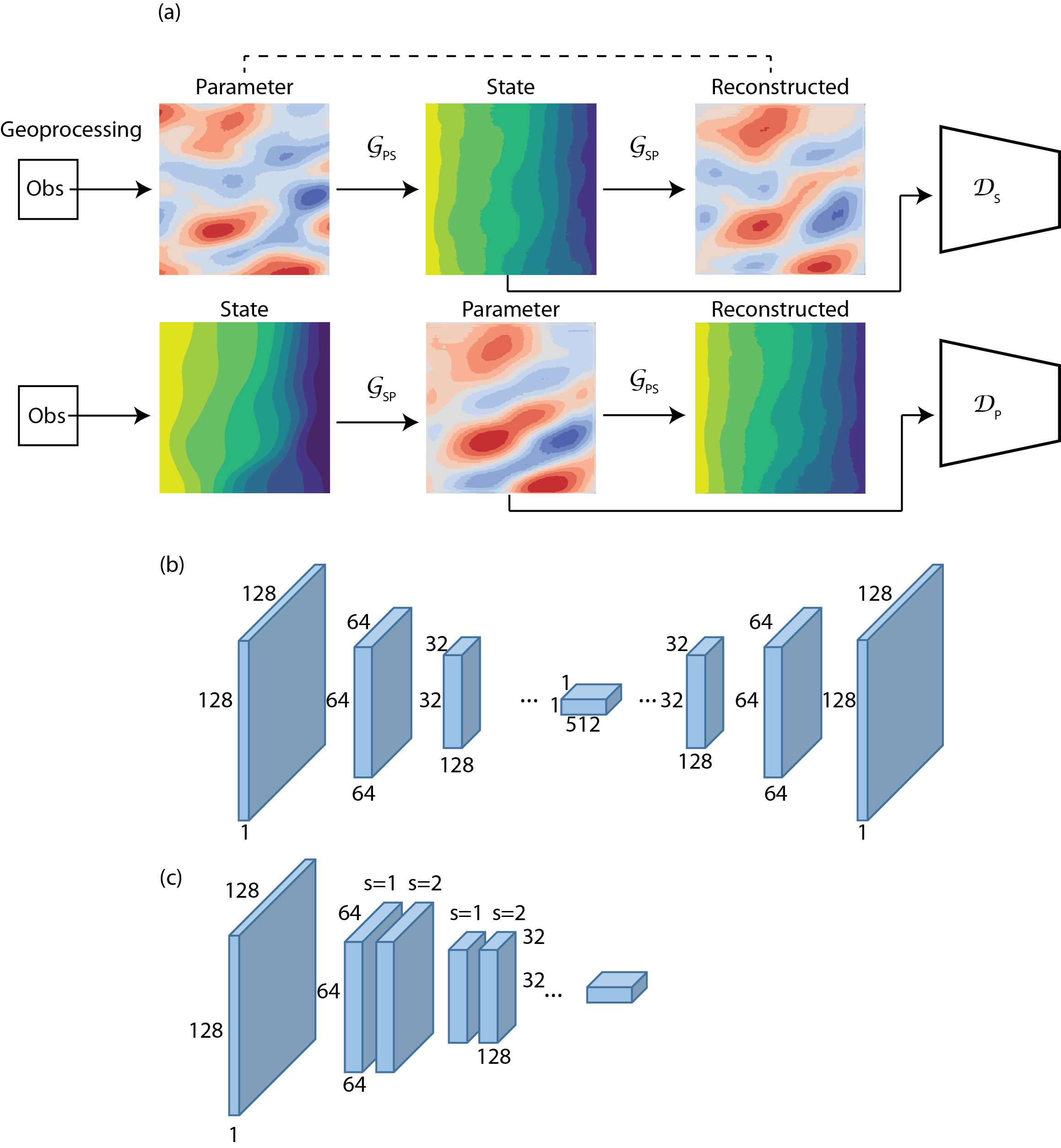}\caption{(a) SPID-GAN consists of two pairs of generators and discriminators
trained together, one pair for identifying the forward parameter-to-state
mapping ($\mathcal{G}_{PS},\mathcal{D}_{S}$) (first row), and the
other for identifying the reverse state-to-parameter mapping ($\mathcal{G}_{SP},\mathcal{D}_{P}$)
(second row); cycle consistency is enforced by minimizing the reconstruction
loss (third column); observations of parameter and state variables
can be fused through pre- and post-processing steps (e.g., kriging
and geostatistical simulation); (b) the design of generator model
follows a downsampling-upsampling pattern using convolutional and
deconvolutional layers to learn features at multiple scales; (c) the
discriminator uses repeating convolutional layers to improve learning
of fine-scale features. \label{fig:1}}
\end{figure}

Figure \ref{fig:1}a illustrates the workflow of SPID-GAN. The first
step shown on the left is related to data preparation, which may be
facilitated by a rich set of tools available from the geostatistics
literature \citep{deutsch1998geostatistical}. For example, if parameter
measurements are available, they may be integrated to generate the
so-called ``conditional realizations'' of the parameter field to
honor prior information. If measurements of state variables are available,
they may be used to generate plausible images of the state field via
kriging. Model state measurements may also be used to select the most
probable parameter fields by choosing the parameter sets that minimize
the differences between GAN-predicted state values and the actual
observed values. Assuming point measurements of parameter and state
variables are available, the SPID-GAN workflow implies an additional
loss term on the generators that is enforced through pre- and post-processing,

\begin{equation}
d_{obs}=\mathbb{E}_{S}\left\Vert \mathcal{G}_{PS}({\bf x})({\bf u}_{S})-{\bf y}({\bf u}_{S})\right\Vert _{2}^{2}+\mathbb{E}_{P}\left\Vert \mathcal{G}_{SP}({\bf y})({\bf u}_{P})-{\bf x}({\bf u}_{P})\right\Vert _{2}^{2},
\end{equation}

\noindent where ${\bf u}_{P}$ and ${\bf u}_{S}$ are locations of
$P$ and $S$ measurements. These different use cases will be illustrated
in the Result section.

In this work, SPID-GAN is implemented by using convolutional neural
networks (CNN), a class of deep feed-forward neural networks specially
designed for image pattern recognition \citep{lecun2015deep}. A brief
introduction of common CNN terminologies is given in Supporting Information
(SI) S1. The two generators share a deep learning neural network design
that is similar to DiscoGAN \citep{kim2017learning}, which includes
a series of convolutional and deconvolutional layers to help uncover
features at multiple scales (Figure \ref{fig:1}b). The input images
have dimensions $128\times128$. The convolutional layers (the downsampling
path) use $4\times4$ kernels and a uniform stride size of 2, while
the deconvolutional layers (the upsampling path) all use a stride
size of 1. The number of filters increases from 64 to 512 for the
convolutional layer stack, while the deconvolutional layers reverse
the number of filters to generate an output image of the same size
as the input. The leaky rectified linear unit function (ReLU) is used
as the activation function for all hidden CNN layers, and the hyperbolic
tangent function $(\tanh)$ is used as the activation function for
the output layer. Instance normalization is applied to hidden layers
to improve the training speed \citep{ulyanov1607instance}. The two
discriminators use the same design as shown in Figure \ref{fig:1}c.
For the hidden layers, the layer configuration is repeated twice using
alternating stride sizes 1 and 2. All codes are written in Python
using the open-source deep learning package, Keras \citep{chollet2015keras}.
The Adam optimization solver \citep{kingma2014adam} is used for training,
with a learning rate 0.0002 and a forgetting factor 0.5. Unless otherwise
noted, the number of epochs used in training is 125 and the batch
size is 10. The computing time taken for each epoch is about 23 s
on a Cloud-based Ubuntu instance running on an Intel Xeon E5-2580
CPU node with GPU acceleration (NVIDIA Tesla K40). 

To quantify the skill of trained generators, the structural similarity
index (SSIM) commonly used in image analysis \citep{wang2004image}
is adopted as a metric. For two sliding windows ${\bf {\bf u}}$ and
${\bf v}$ of dimensions $n_{p}\times n_{p}$, SSIM is defined as
\begin{equation}
SSIM({\bf u},{\bf v})=\frac{(2\mu_{{\bf u}}\mu_{{\bf v}}+c_{1})(2\sigma_{{\bf uv}}+c_{2})}{(\mu_{{\bf u}}^{2}+\mu_{{\bf v}}^{2}+c_{1})(\sigma_{{\bf u}}^{2}+\sigma_{{\bf v}}^{2}+c_{2})},
\end{equation}

\noindent where ${\bf u}$ and ${\bf v}$ represent two patches from
the simulated image (using GAN) and testing image (from numerical
model), respectively, $\mu$ represents the mean, $\sigma^{2}$ represents
the variance, and $c_{1}$ (0.01) and $c_{2}$ (0.03) are small constants
used to stabilize the denominator. The mean value of SSIM, averaged
over all sliding windows, ranges from -1 to 1, with 1 being attainable
when two images are identical. The size of the sliding window used
in this study is $n_{p}=7$. 

\subsection{Groundwater flow }

To demonstrate the usefulness of SPID-GAN for learning the bidirectional
mappings, I consider groundwater flow in a spatially heterogeneous
aquifer, which is a representative geoscience problem and has been
studied extensively. The governing equation is given by the following
PDE
\begin{equation}
S_{s}({\bf z})\frac{\partial h({\bf z},t)}{\partial t}=\nabla\cdot\left(K({\bf z})\nabla h({\bf z},t)\right)+q_{w}({\bf z},t),\label{eq:9}
\end{equation}

\noindent where $h$ {[}L{]} is hydraulic head, $S_{s}$ {[}1/L{]}
is specific storage, $K$ {[}L/T{]} is hydraulic conductivity, ${\bf z}$
denotes spatial coordinates, $t$ is time, and $q_{w}$ {[}L\textsuperscript{3}/T{]}
is the source/sink term. For the purpose of illustration, the state
variable is $h$ and the parameter is $K$, both are spatially distributed
variables. All other quantities are assumed deterministic. The lateral
dimensions of the aquifer are $1280\textrm{ m}\times1280\textrm{ m}$,
and the domain is uniformly discretized into $10\textrm{ m\ensuremath{\times}10}\textrm{ m}$
grid blocks. The thickness of the aquifer is 20 m. The goal of SPID-GAN
is to train two generators simultaneously to approximate the physical
model specified in Eq. (\ref{eq:9}). Two different problem settings
are considered, single- and multimodal parameter distributions. 

\section{Results}

\subsection{Single-modal parameter distribution}

In the first problem, $K$ is assumed to be a random field following
log-normal distribution. The mean and standard deviation of $\log K$
are $2\times10^{-4}$ and 1.0. The variogram model of $\log K$ is
Gaussian with max and min ranges of 500 and 100 m, respectively. $S_{s}$
is deterministic with a value of $2.5\times10^{-6}$ $\mathrm{m}^{-1}$.
In the baseline case, constant-head (Dirichlet) boundary conditions
are imposed on both the west (21 m) and east (10 m) sides of the aquifer.
Four pumping wells are put in grid blocks (25, 25), (25,106), (106,
106), and (106,25), with pumping rates of 10, 10, 50, and 20 m\textsuperscript{3}/day,
respectively. Stochastic realizations of $\log K$ are generated by
using the sequential Gaussian simulator ($\texttt{sgsim}$), available
from the open-source geostatistical package SGeMS \citep{remy2009applied}.
The flow field is first run to the steady state, followed by a transient
simulation period of 3600 s. For training and validation, the corresponding
head distributions are obtained by using the open-source groundwater
flow solver MODFLOW via its Python wrapper, $\texttt{flopy}$ \citep{bakker2016scripting}.
The computing time for each forward simulation is 0.04 s. Such geostatistical
processing steps have been broadly used in previous studies, such
as ensemble-based data assimilation\citep{chen2006data,franssen2009comparison,sun2009comparison},
hydraulic tomography \citep{lee2014large}, and surrogate modeling
and uncertainty quantification \citep{li2007probabilistic,nowak2010bayesian}.
Training of the SPID-GAN is done using 400 pairs of $\textrm{log}K$
(parameter domain) and $h$ (state domain) fields, each having dimensions
of $128\times128$.

\begin{figure}
\noindent \centering{}\includegraphics[width=6in]{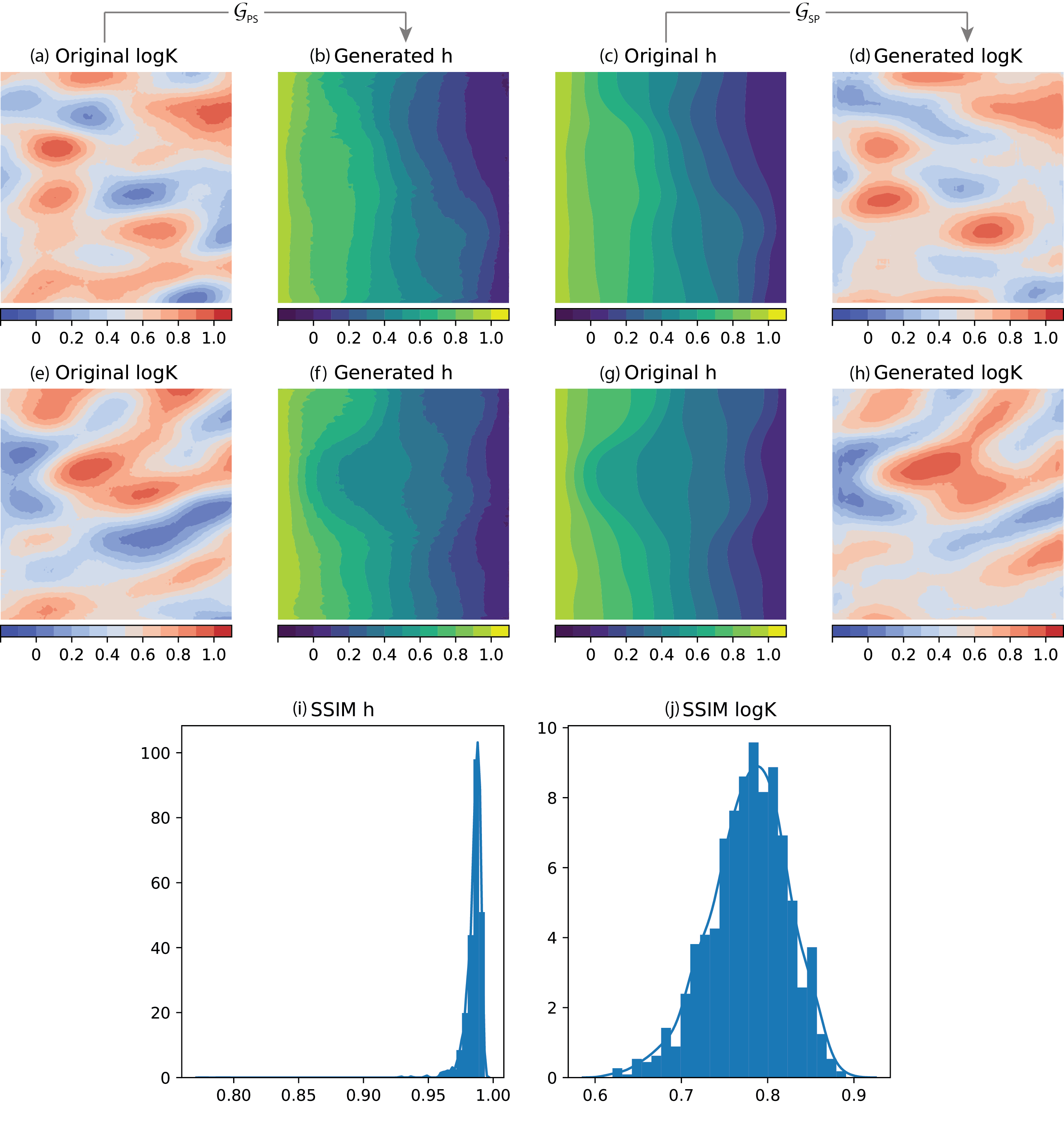}\caption{SPID-GAN results for two random realizations (a-d and e-h): (a) and
(e) are original $\log K$ realizations generated using $\texttt{sgsim}$,
(c) and (d) are original head images simulated using MODFLOW; head
fields (b) and (f) are generated using the trained forward generator
$\mathcal{G}_{PS}$, while parameter fields (d) and (h) are generated
by the trained reverse generator $\mathcal{G}_{SP}$. All contours
are normalized for visualization purposes. Subplots (i) and (j) show
histograms of structural similarity indices (SSIM) calculated on 1000
test samples not used during training. The mean SSIM of $\mathcal{G}_{PS}$
is 0.98 and the mean SSIM of $\mathcal{G}_{SP}$ is 0.78. \label{fig:2}}
\end{figure}

Figures \ref{fig:2}a-d and e-h show examples resulting from two test
realizations not included in training. The left two images of each
row show the input $\log K$ field and the corresponding head field
generated using the trained forward generator $\mathcal{G}_{PS}$,
while the right two images of each row show the input state (head)
and corresponding $\log K$ field predicted using the trained reverse
generator $\mathcal{G}_{SP}$. For this base case, SPID-GAN captures
the spatial patterns in head distributions well, on the basis of visual
comparisons between Figures \ref{fig:2}b and \ref{fig:2}c, and between
Figures \ref{fig:2}f and \ref{fig:2}g. Results of testing using
1000 test realizations of $\log K$ give an ensemble mean SSIM value
of about 0.98 (Figure \ref{fig:2}i). In comparison, by visual examinations
of Figures \ref{fig:2}a and \ref{fig:2}d, and Figures \ref{fig:2}e
and \ref{fig:2}h, it can be seen that the trained reverse mapping
$\mathcal{G}_{SP}$ captures the dominant spatial patterns in the
original $\log K$ images, but tends to underestimate some local maxima
or minima. The ensemble mean SSIM obtained from testing on 1000 $h$
field samples (not used in training) is 0.78. 

In general, learning a high-dimensional reverse mapping is a significantly
more challenging problem to solve, depending on not only the design
of the GAN algorithm, but also the experimental design and quality
of training samples. In this case, the learning of reverse mapping
would benefit from any conditions (e.g., boundary conditions and forcings)
that can help improve the information content of head fields and the
uniqueness of cross mappings. In an experimental design, if the head
field is not sensitive to certain parts of the parameter field, SPID-GAN
may give ambiguous inverse solutions in those areas. To elaborate
this latter point, in SI S2 the number of pumping wells is reduced
from four to two (Figure S2), and then to none (Figure S3). As a result,
the parameter fields in those cases become less identifiable\textemdash more
artifacts start to appear in reversely mapped $\log K$ fields in
those cases, especially in the central part of the domain that is
less stimulated than the parts near the west and east boundaries.
Nevertheless, the overall SPID-GAN performance stays relatively robust,
as can be seen by the SSIM statistics in the respective plots. 

In addition to different boundary/forcing conditions, the effect of
training sample quantity on SPID-GAN performance is also investigated.
The results, shown in SI S3, suggest that the GAN is relatively robust
when the number of training samples varies, indicating the capability
of GAN to learn dominant cross-domain patterns.

\begin{figure}
\centering{}\includegraphics[height=8in]{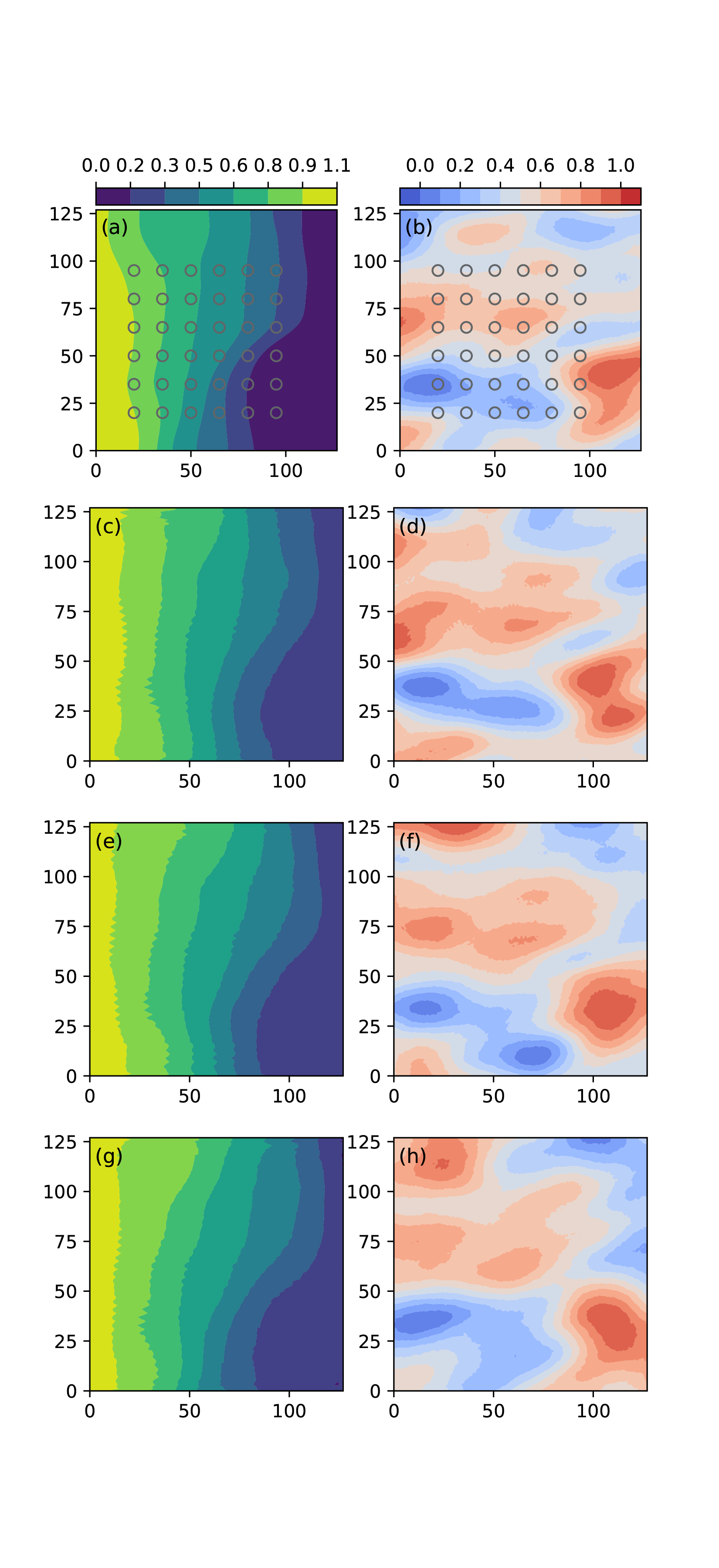}\caption{Illustration of the use of prior information in SPID-GAN: (a) and
(b) show the ``true'' $\log K$ and head fields, which are sampled
only at monitoring locations (open circles); (c), (e), (g) show the
top three $\log K$ realizations identified as the ``closest'' to
the true head field, as measured using MSE between simulated and observed
head values; (d), (f), (h) show the corresponding head fields, which
resemble the true head field. \label{fig:3}}
\end{figure}

GANs operate with images, while the traditional workflow in hydrogeology
typically involves point measurements. This next example demonstrates
that the two workflows are actually complementary under the new SPID-GAN
workflow proposed in Figure \ref{fig:1}a. Figures \ref{fig:3}a,
b show the ``true'' $\log K$ and head fields, which are sampled
only at limited locations. For simplicity, it is assumed that point
observations of $K$ and $h$ are collected from the same monitoring
network (open circles in Figures \ref{fig:3}a and \ref{fig:3}b,
a total of 36 conditioning points), although they may well be different.
First, 1000 $\log K$ conditional realizations are generated using
$\texttt{sgsim}$ that honor the prior information on $K$. For each
conditional $\log K$ realization, the trained forward generator $\mathcal{G}_{AB}$
from the base case is used to predict a head field, which is then
sampled at monitoring locations to calculate the MSE between the simulated
and actual head observations. The resulting MSE values are sorted
in an ascending order. Figures \ref{fig:3}c,e,g show the top 3 head
fields that have the minimal MSE, while the corresponding $\log K$
fields are plotted in Figures \ref{fig:3}d,f,h. The identified fields
show strong resemblance to the synthetic truth, especially where conditional
information is available (i.e., comparing SPID-GAN results to the
synthetic truth at the monitoring locations). Similarly, the workflow
presented here can be used to form an ensemble of models to perform
the DL-based, uncertainty quantification in the sense of the recently
introduced data space inversion, in which the goal of modeling is
not calibration in the traditional sense, but to establish a data-driven
statistical relationship between the observed and forecast variables
and to quantify the predictive uncertainty of the forecast variables,
by using an ensemble of uncalibrated prior models \citep{satija2015direct,sun2017production,jeong2018learning}.

\subsection{Bimodal parameter distribution}

In the second problem setting, the feasibility of using SPID-GAN for
learning multimodal distributions is investigated. The aquifer is
assumed to consist of two hydrofacies, a permeable channel facies
and a background matrix, making the distribution of $K$ bimodal.
Identification of facies shapes is a representative and yet challenging
inversion problem that has also been studied extensively in the literature
\citep{liu2005ensemble,sun2009sequential,zhou2011approach}. In this
example, the channel facies has a $K$ value of $2\times10^{-4}$
m/s and an $S_{s}$ value of $1\times10^{-6}$ $\mathrm{m}^{-1}$,
while the matrix has a $K$ value of $1\times10^{-8}$ m/s and an
$S_{s}$ value of $1\times10^{-7}$ $\mathrm{m}^{-1}$. The facies
realizations are generated using $\texttt{snesim}$, which is a multipoint
geostatistical simulator also available from SGeMS \citep{remy2009applied}.
Constant heads of 11 and 10 m are imposed on the west and east boundaries,
and the transient simulation period is 5400 s. All other settings
are the same as used in the previous two examples. 

\begin{figure}
\centering{}\includegraphics[width=6in]{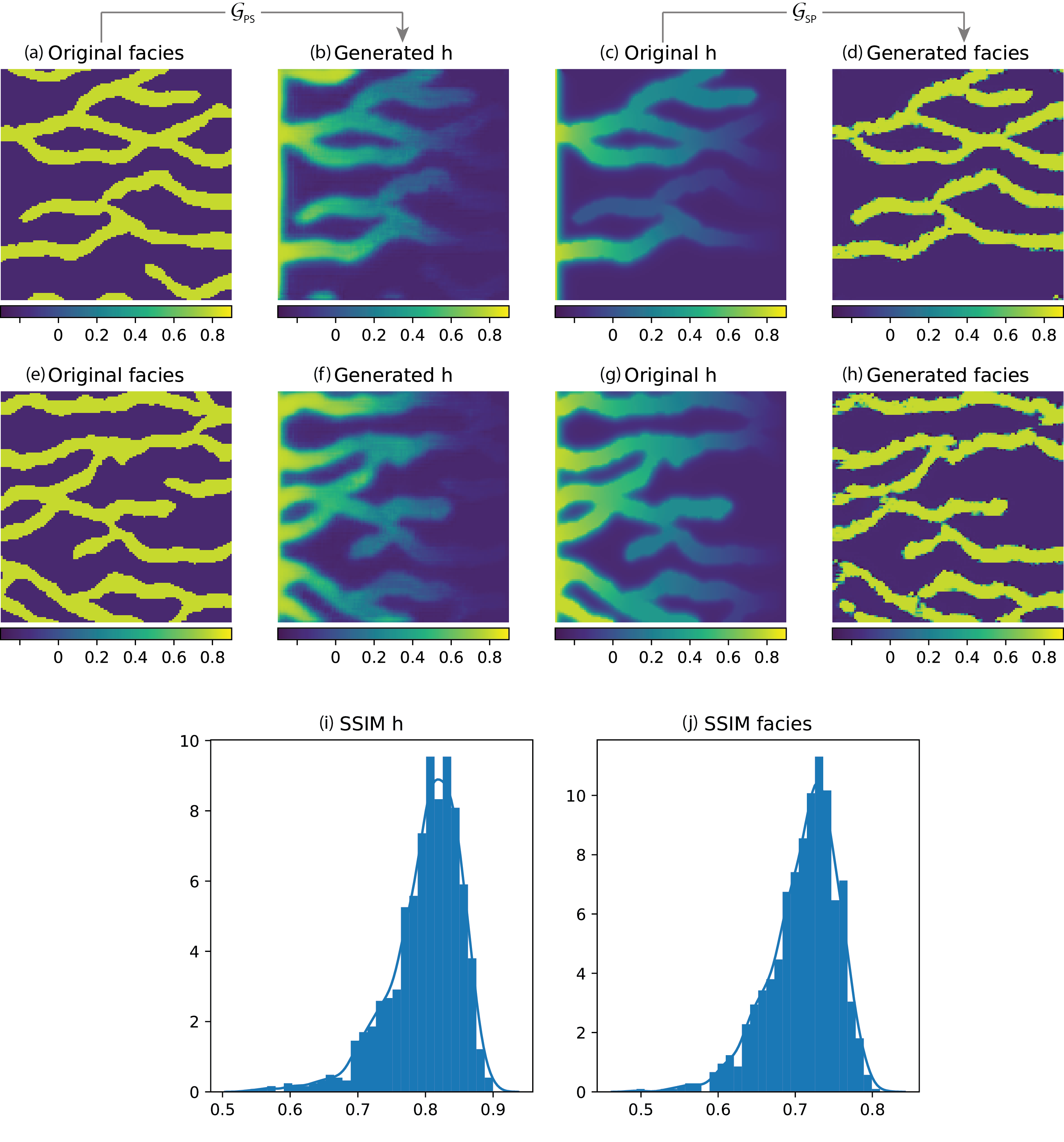}\caption{Illustration of the use of SPID-GAN to identify bidirectional mappings
for a bimodal parameter distribution: subplot (a)-(d) show results
from a test realization; subplot (e)-(h) show results from another
test realization. The SSIM histograms obtained using 1000 test realizations
are shown in subplots (i)-(j). The mean SSIM of the forward generator
is 0.8, and the mean SSIM of the reverse generator is 0.71. \label{fig:4}}
\end{figure}

Figures \ref{fig:4}a-d and e-h show example results from two test
realizations. The forward generator predicts the connected flow paths
well, including the fading color pattern in head distribution from
the west boundary to the east boundary (comparing Figures \ref{fig:4}b
and \ref{fig:4}c, and Figures \ref{fig:4}f and \ref{fig:4}g). The
reverse mapping from the head distribution identifies the well-connected
backbone of the flow network (comparing Figure \ref{fig:4}a with
\ref{fig:4}d, and Figure \ref{fig:4}e with \ref{fig:4}h), but tends
to omit small channel segments that are not connected to the main
flow pathways. This is because the information of disconnected segments
is not discoverable from head distributions. In other words, the head
measurements are not sensitive to the parameter information in disconnected
channel segments. From a practical standpoint, the connected features
are more important to identify for risk assessment purposes. Finally,
Figures \ref{fig:4}(i) and (j) show the SSIM statistics on running
1000 pairs of randomly generated test samples. Again SPID-GAN gives
a reasonable performance in this bimodal distribution case, with an
ensemble mean SSIM value of 0.8 for forward mapping ($K$ to $h$)
and 0.71 for reverse mapping ($h$ to $K$).

\section{Summary and Conclusions}

The recently introduced generative adversarial networks (GANs) have
shown strong performance in image-to-image translation. Questions
remain about whether GANs can be used to learn the deep causal structures
embedded in physical models. In this study, I show that the usefulness
of GANs goes beyond the simple picture mapping and demonstrate the
feasibility of using a state-parameter identification GAN (SPID-GAN)
to approximate bidirectional mappings between the parameter space
and state space of PDEs. Through a series of representative problems
selected from groundwater modeling, I show that SPID-GAN does a satisfactory
job (with a reasonable amount of computing time) in linking physical
parameters to model states, and is able to capture the complex spatial
patterns in parameter and state distributions that are otherwise challenging
to obtain. The need for identifying forward and reverse mappings is
ubiquitous in geophysical research. Thus, findings of this study may
have important implications for many similar tasks, such as geostatistical
simulation, history matching, uncertainty quantification, surrogate
modeling, optimal design of experiments, or any problem that requires
sampling from a high-dimensional distribution in a computationally
efficient manner. 

\section*{Acknowledgement}

The work was partly supported by the U.S. Department of Energy, National
Energy Technology Laboratory (NETL) under grant numbers DE-FE0026515
and DE-FE0031544. The author is grateful to the handling editor, Prof.
Valeriy Ivanov, and two anonymous reviewers for their constructive
comments. All the data used are listed in the references or archived
in repository, \url{https://utexas.box.com/s/uq74paot1a2ns3v9hbll01ifev408uus}.

\bibliographystyle{agufull08}

\pagebreak{}

\textbf{Figure Captions}

1. (a) SPID-GAN consists of two pairs of generators and discriminators
trained together, one pair for identifying the forward parameter-to-state
mapping ($\mathcal{G}_{PS},\mathcal{D}_{S}$) (first row), and the
other for identifying the reverse state-to-parameter mapping ($\mathcal{G}_{SP},\mathcal{D}_{P}$)
(second row); cycle consistency is enforced by minimizing the reconstruction
loss (third column); observations of parameter and state variables
can be fused through pre- and post-processing steps (e.g., kriging
and geostatistical simulation); (b) the design of generator model
follows a downsampling-upsampling pattern using convolutional and
deconvolutional layers to learn features at multiple scales; (c) the
discriminator uses repeating convolutional layers to improve learning
of fine-scale features.

2. SPID-GAN results for two random realizations (a-d and e-h): (a)
and (e) are original $\log K$ realizations generated using $\texttt{sgsim}$,
(c) and (d) are original head images simulated using MODFLOW; head
fields (b) and (f) are generated using the trained forward generator
$\mathcal{G}_{PS}$, while parameter fields (d) and (h) are generated
by the trained reverse generator $\mathcal{G}_{SP}$. All contours
are normalized for visualization purposes. Subplots (i) and (j) show
histograms of structural similarity indices (SSIM) calculated on 1000
test samples not used during training. The mean SSIM of $\mathcal{G}_{PS}$
is 0.98 and the mean SSIM of $\mathcal{G}_{SP}$ is 0.78. 

3. Illustration of the use of prior information in SPID-GAN: (a) and
(b) show the ``true'' $\log K$ and head fields, which are sampled
only at monitoring locations (open circles); (c), (e), (g) show the
top three $\log K$ realizations identified as the ``closest'' to
the true head field, as measured using MSE between simulated and observed
head values; (d), (f), (h) show the corresponding head fields, which
resemble the true head field. 

4. Illustration of the use of SPID-GAN to identify bidirectional mappings
for a bimodal parameter distribution: subplot (a)-(d) show results
from a test realization; subplot (e)-(h) show results from another
test realization. The SSIM histograms obtained using 1000 test realizations
are shown in subplots (i)-(j). The mean SSIM of the forward generator
is 0.8, and the mean SSIM of the reverse generator is 0.71.

\newpage{}


\end{document}